\documentclass[twocolumn,preprintnumbers,amsmath,amssymb,prl]{revtex4}
\usepackage{graphicx}% Include figure files
\usepackage{dcolumn}% Align table columns on decimal point
\usepackage{bm}% bold math

\begin{document}

\preprint{Nature Materials {\bf 3}, 659 (2004)}

\title{Looking for Design in Materials Design}
\author{Mark E. Eberhart}
\affiliation{
Department of Chemistry \& Geochemistry\\
Colorado School of Mines\\
Golden, CO 80401}

\author{Dennis P. Clougherty}
\affiliation{
Department of Physics\\
University of Vermont\\
Burlington, VT 05405-0125}

\date{\today}

\maketitle

%\section{introduction}

The emerging discipline of computational materials design (CMD) seeks to speed materials development by using computers to calculate the properties of substances that have yet to be made in the laboratory.  In this way, computation could one day supplant time-consuming experimental empiricism, greatly accelerating the pace of technological advancement \cite{materials}. Over the past twenty years, the potential of CMD has attracted hundreds of millions of dollars in investment.  In some ways, this investment has paid dividends by providing the ability to perform computer simulations of ever-greater sophistication. However, these new capabilities have had less influence on materials development than its supporters might have predicted; other than a few small molecules \cite{design}, not a single material has been computer-designed.  We suggest that the path towards achieving the ultimate goals of this still immature science is in need of a midcourse adjustment. 

Both scientists and society-at-large have a stake in seeing CMD reach its full potential.  Such a science could dramatically and radically alter the way technologies develop, consequently liberating the pace of technological advance from its current dependence on the unpredictability of trial-and-error experiments.  For millennia, technologies were born only when a new material with novel properties became available. During the past century, science developed a limited capability to design materials, but we are still too dependent on serendipity.   In just the past twenty years, researchers have stumbled upon high-temperature superconductors, conducting polymers, quasicrystals, and nanotubes-all possessing properties that, before their discovery, conventional wisdom deemed impossible.  Now, these materials are being explored as the foundation of a host of new twenty-first-century technologies. But a full science of materials design would enable materials such as these to be designed in response to the needs of technology-- that is, the path should lead from properties to materials --instead of the reverse. 

Our primitive abilities to design materials have developed concurrently with a fundamental understanding of the laws of physics and chemistry that govern the properties of materials.  Although these laws were fully elucidated more than seventy-five years ago, they gave rise to equations that--for any real-world material--were much too difficult to solve. Yet, two advances have now made numerical solutions practicable: (1) the relentless progress in digital computers that have boosted processing speed by more than five orders of magnitude over the past twenty years; and (2) the improvements in computational methods. Now, there is a growing consensus that a general science of CMD will one day be achievable. 

In the view of many, the structure for this new science will rest upon the foundation provided by solving the equations of quantum mechanics, which provide the thermodynamic rationale to account for all materials properties. Accordingly, beginning in the mid 1980s and continuing to this day, U.S. agencies, including the Defense Advanced Research Projects Agency (DARPA), the National Science Foundation (NSF), and the Department of Energy (DOE), aggressively funded interdisciplinary programs designed to improve or extend computational quantum mechanics software. The idea was to combine the experience of computer scientists, materials scientists, and computational physicists to simulate and predict materials properties.

The investments of these funding agencies spurred innovations in density functional theory (DFT), which provided a way to circumvent many of the difficulties of the underlying quantum many-body problem, with dramatic results. Where twenty years ago, only idealized systems could be modeled, today the emphasis is on the simulation of ``real materials.''  Most notable is the use of DFT to calculate the thermodynamic internal energies, charge densities, and wave functions of property-determining defects in crystalline materials, e.g. metals, semiconductors, and ceramics.  However, in only a few cases have these computations played a part in the development of a new substance. Moreover, the breakthrough in these instances had not come by using the most advanced simulation techniques possible--those that calculate thermodynamic quantities most accurately. It resulted from integrating computation with materials design in the same way computation has been integrated with established forms of design.  

As an example, computational fluid dynamics often supplants wind tunnel testing in the design of a number of commercial products, including automobiles, airplanes, bicycles, sailboats and skis. The drag coefficient, a measure of the resistance to motion of an object in a fluid, can be calculated; however, its value does not give a designer insight into the causes of drag. The structure of the flow field around the airplane wing or ski is of greater value because turbulence in the flow field is an indicator of drag. Rather than calculating the drag coefficient, observing this field in a wind tunnel or working directly with the computationally determined flow field has the advantage of pointing the designer directly to the source of the turbulence. Identifying the source of the turbulence provides the most efficient way to decrease drag. 

The way turbulence controls the efficiency of an object moving through a fluid is an example of a structure-property relationship.  These relationships form the basis for all design. Turbulence is a structure that gives rise to drag, a property. Our ability to design materials is limited by incomplete knowledge of all the structure-property relationships governing the behaviours of materials. Without this knowledge, materials development must proceed by trial and error, which is both costly and time-consuming, with some materials requiring decades of research and development before technological applications can be found.

Given the importance of structure-property relationships in materials design, it is understandable that uncovering new relationships is an important focus of materials research. Unfortunately, no standard methodology exists to guide researchers in this discovery. Yet, history suggests that new structure-property relationships emerge as our capacity to search for patterns at different scales of resolution improves.   Cyril Stanley Smith observed \cite{smith} that structure is best considered as a hierarchy, with each of its levels characterized by a different length scale. Macrostructure sits at the apex of the hierarchy, where the distinguishing length scale is that resolvable by the unaided eye. At the base, with a distinguishing length scale on the order of less than an angstrom, sits the charge density.  Because our ability to inspect structure at this scale is limited, sub-angstrom scale structure-property relationships are for the most part unknown to materials science.  But our ability to inspect structure at the sub-angstr{\"o}m scale is limited, and structure-property relationships at this level are for the most part unknown to materials science.

Fortunately, advances in quantum mechanical computation have provided new capabilities.  It is now possible to calculate the charge density of real materials and to search for relationships between this density and properties.  At the same time, it is also often possible to directly calculate the thermodynamic properties of materials of a known structure.  It is the latter capability that has been most exploited and advancedÑseeking ever more accurate calculation.  This is true even though quantum mechanical calculations have proved most useful as design tools when used to uncover new structure-property relationships.  Nobel-prize -winning chemist Roald Hoffmann made this same point regarding the quantum mechanical theories used to design molecules. Hoffmann and his colleague Robert Woodward showed in 1969 \cite{w&h} that all chemical reactions are controlled by the pattern of nodes in specific electronic orbitals of the reacting molecules. With the discovery of this structure-property relationship, ``very poor quality but very useful'' \cite{nyt} quantum calculations became a standard tool of the synthetic chemist. What makes these calculations ``poor quality'' is that they do not accurately calculate the thermodynamic quantities; what makes them ``very useful'' is that they can predict nodal patterns. Thus, quantum calculations are beneficial in creating new chemical compounds when used in the context of this particular structure-property relationship, where nodal patterns control chemical reactivity. 

Hoffman's suggestion-- that the accuracy of quantum calculations can and should be sacrificed for the sake of predicting structure-property relationships-- is hard to swallow for many researchers, who are working to develop ever more sophisticated computational techniques. If simulations can be conducted accurately and sufficiently rapidly, many workers argue, then computers can search for materials with properties optimized for the intended application without relying on the unpredictable discovery of new relationships between structure and properties. This vision is certainly attractive, and it may actually pay off; however, this is not materials design.  For no matter how fast the computer, if it must search for the optimum properties among an infinite number of materials, it will still require infinite time to perform the search. Design, on the other hand, begins with a set of desired properties, and using structure-property relationships, identifies a finite set of candidate materials that are likely to possess these properties --in accord with Hoffman's observation regarding the utility of quantum calculations in molecular design. These materials may then be studied experimentally.  

\begin{figure}
\includegraphics[width=3in]{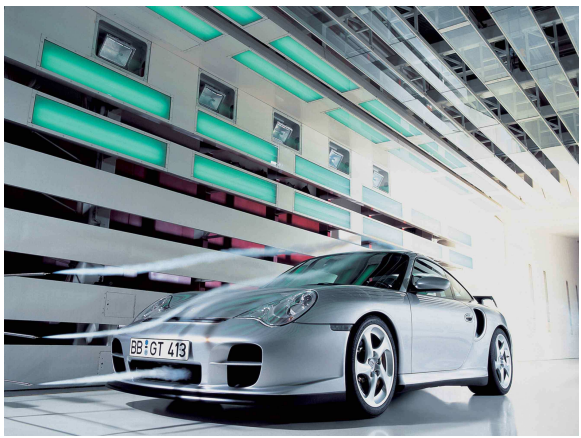}
\caption{\label{car} 2002 Porsche 911 GT2 coupe in a wind tunnel.  The lamellar flow of the smoke along the hood is a visual indication of a low aerodynamic drag. (Image courtesy of Porsche Cars.) }
\end{figure}

There are parallels between computational fluid dynamics and the use of quantum mechanics in materials design that may serve to guide the midcourse adjustment of CMD.  The utility of computational fluid dynamics arises because drag is a functional of the flow field.  As a result, we are able to qualitatively associate the topological and geometrical properties of this field with increasing drag.  A quick survey of the popular press--from Sports Illustrated, showing an athlete in a wind tunnel, to Car and Driver, depicting the computed flow across the latest sport car--serves to emphasize that topological and geometrical characteristics of the flow, vortices and turbulence, have become icons for drag. (See Fig.~\ref{car}.) Similarly, by DFT, a materialÕs thermodynamic energy is a functional of its charge density, which intimates that there might be relationships between the topological and geometrical properties of the charge density and many materials properties of interest to designers.  Yet, there has been little effort to identify these relationships. For example, often when developing materials, it is desirable to alter the stability of a particular substance. And though this stability is determined by the geometry of the charge density, we know of no relationship that qualitatively associates characteristics of the charge density with stability.  If there were such a relationship, the orbital theory of atomic bonding might allow us to predict how various atomic impurities would alter the charge density, making it possible to design for stability.

It is regrettable that computational materials science has been so unbalanced, with the dominant endeavour within the field directed towards improving our ability to calculate the properties of a given structure. If this discipline is to realize its full potential, we must seek a more equitable distribution of effort and resources and turn our attention to identifying relationships between materials properties and quantum mechanical structure--the charge density.  Ultimately, we must be able to specify the charge density that is likely to give rise to a set of desired properties.  Unfortunately, materials scientists, and the agencies that support their research, have been slow to recognize the importance of these capabilities to the ultimate success of this emerging discipline.  This situation must change if the tremendous advances achieved in materials modeling and simulation are to play a significant role in materials design.

We thank the Air Force Office of Scientific Research and the Defense Advanced Research Projects Agency for the support of our research.

\end{document}